\documentclass[prl,twocolumn,nofootinbib,showkeys,citeautoscript]{revtex4}
\pdfoutput=1
\usepackage{graphicx}
\usepackage{epstopdf}
\usepackage{amsmath}
\usepackage{hyperref}
\usepackage{cleveref}
\crefname{equation}{Eq.}{Eqs.}
\crefname{figure}{Fig.}{Figs.}
\crefname{table}{Table}{Tables}
\newcommand{\crefrangeconjunction}{--}
\usepackage{dcolumn}
\usepackage{fancybox}
\usepackage{bm}
\usepackage{multirow}
\usepackage[usenames,dvipsnames]{color}
\usepackage{array}
\usepackage{rotating}
\usepackage{ulem,fancyvrb}
\usepackage{latexsym}
\usepackage{amsfonts}
\usepackage{xcolor}
\usepackage{longtable}
\usepackage{amssymb}
\usepackage{xspace}
\def\tana2{\tan^2\alpha_H}
\def\tanb2{\tan^2\beta}

\def\m0pr{m_0'}

\def\Mz2{M_Z^2}

\def\met100{\slashed{E}_T\geq 100~{\rm GeV}}

\newcommand{\beqn}{\begin{eqnarray}}
\newcommand{\eeqn}{\end{eqnarray}}
\newcommand{\be}{\begin{equation}}
\newcommand{\ee}{\end{equation}}
\def \cha{\tilde{\chi}^{\pm}_1}

\newcommand{\na}{\ensuremath{\tilde{\chi}^{0}_1}}

\def \nb{\tilde{\chi}^{0}_2}
\def \nc{\tilde{\chi}^{0}_3}
\def \nd{\tilde{\chi}^{0}_4}
\def \n34{\tilde{\chi}^{0}_{3,4}}
\newcommand{\g}{\ensuremath{\tilde{g}}}
\newcommand{\q}{\ensuremath{\tilde{q}}}

\def \ta{\tilde{t}_1}

\def \ba{\tilde{b}_1}

\def \mhf{m_{1/2}}

\newcommand{\mh}{\ensuremath{m_{h^0}}}

\newcommand{\etal}{{\it et al.}, }

\def\ra{\rightarrow}
\newcommand{\bs}{\ensuremath{B_s^0}}

\newcommand{\mm}{\ensuremath{\mu^{+}\mu^{-}}}

\newcommand{\bsmm}{\ensuremath{\bs\ra\mm}}

\newcommand{\brbsmm}{\ensuremath{\mathcal{B}r(\bsmm)}}

\newcommand{\msb}{\ensuremath{{\overline{\mathrm{MS}}}}}
\newcommand{\brbsg}{\ensuremath{\mathcal{B}r(\bar{B}\to X_s\gamma)}}
\newcommand{\sta}{\ensuremath{\tilde{\tau}_1}}

\def \smr{\tilde{\mu}_R}
\def \ser{\tilde{e}_R~}

\DeclareMathOperator{\sgn}{sgn}

\newcommand{\TeV}{\ensuremath{\,\text{Te\kern -0.1em V}}\xspace}
\newcommand{\GeV}{\ensuremath{\,\text{Ge\kern -0.1em V}}\xspace}
\newcommand{\MeV}{\ensuremath{\,\text{Me\kern -0.1em V}}\xspace}

\newcommand{\tev}{\ensuremath{\text{Te\kern -0.1em V}}\xspace}

\allowdisplaybreaks[1]

\begin{document}

\title{Implications of the Higgs Boson Discovery for mSUGRA}
\author{Sujeet~Akula}  
\affiliation{Department of Physics, Northeastern University,
 Boston, MA 02115, USA}

\author{Pran~Nath} 
\affiliation{Department of Physics, Northeastern University,
 Boston, MA 02115, USA}

\author{Gregory~Peim}
\affiliation{Department of Physics, Northeastern University,
 Boston, MA 02115, USA}

 \begin{abstract}
A Bayesian analysis is carried  out to identify the consistent regions of the mSUGRA parameter space, where the newly-discovered Higgs boson's mass is used as a constraint, along with other experimental constraints. It is found that $m_{1/2}$ can lie in the sub-TeV region, $A_0/m_0$ is mostly confined to a narrow strip with $|A_0/m_0| \leq 1$, while $m_0$ is typically a TeV or larger.  Further, the Bayesian analysis is used to set 95\% CL lower bounds on sparticle masses.  Additionally, it is shown that the spin independent neutralino-proton cross section lies just beyond the reach of the current sensitivity but within the projected sensitivity of the SuperCDMS-1T and XENON-1T experiments, which explains why dark matter has thus far not been detected. The light sparticle spectrum relevant for the discovery of supersymmetry at the LHC are seen to be the gluino, the chargino and the stop with the gluino and the chargino as the most likely candidates. 
\end{abstract}

\maketitle

{\it Introduction:}
The most recent search at the LHC~\cite{July4,HiggsDiscoveryPapers} for the Higgs boson~\cite{HiggsBoson} 
with the combined 7\TeV and 8\TeV data indicates a  signal for the Higgs boson with mass $125.3\pm 0.4\,(\text{stat.})\pm 0.5\,(\text{syst.})\GeV$
for CMS with a local significance of $5.0\,\sigma$ and with mass $126.0\pm 0.4\,(\text{stat.})\pm 0.4\,(\text{syst.})\GeV$ for ATLAS with local significance of $5.9\,\sigma$.
 As is well known the Higgs boson mass at the tree level lies below the $Z^0$ boson 
mass, but it can be made larger by inclusion of loop corrections. 
However, in supergravity grand unification~\cite{can} there is another upper limit, i.e., of about $130\GeV$
due to the constraint of radiative breaking of the electroweak symmetry (for a review see~\cite{IbanezRoss})
as well as other experimental constraints (for a recent analysis see~\cite{Akula:2011aa,Arbey:2012dq}).  
The correction to the Higgs boson mass is given by~\cite{1loop} (for a review see~\cite{review})
\be
\Delta \mh^2\simeq  \frac{3m_t^4}{2\pi^2 v^2} \ln \frac{M_{\rm S}^2}{m_t^2} 
+ \frac{3 m_t^4}{2 \pi^2 v^2}  \left(\frac{X_t^2}{M_{\rm S}^2} - \frac{X_t^4}{12 M_{\rm S}^4}\right)~,
\label{tloop}
\ee
where  $X_t\equiv A_t - \mu \cot\beta$, where $A_t$ is the $A_0$ parameter run down to the weak scale (see \cref{soft}), 
$v=246\GeV$, and $M_{\rm S}$ is an average stop mass.
The loop correction in \cref{tloop} is maximized when $X_t\sim \sqrt{6} M_S$. 
There are also additional loop corrections from, e.g., the $b$-quark sector
as well as from higher loops.
The early searches at the LHC-7 gave a possible hint of the Higgs boson in the 
mass range $\sim (117-129)\GeV$~\cite{lhc_7tev} and the combined Tevatron analysis reported
an excess between $(115-140)\GeV$~\cite{tevatron}.  These findings have led to significant 
activity~\cite{Akula:2011aa,higgs_7tev1,higgs_7tev2} to investigate the implications
of the results for supersymmetry. 

{\it Implications for mSUGRA:}
We note that the scale $M_S$ in \cref{tloop} which is determined by the soft parameters
depends sensitively on the Higgs mass.  In the analysis we use the Higgs boson mass constraint
within the Bayesian statistical framework to estimate the soft parameters of mSUGRA (sometimes 
referred to as CMSSM) which are given by~\cite{can}
\be
 m_0, m_{1/2},  A_0, \tan\beta, \sgn(\mu) \label{soft} 
\ee
where $m_0$ is the universal scalar mass,  $m_{1/2}$ is the universal gaugino mass, $A_0$ is the trilinear 
couplings and $\tan\beta$ is the ratio of the two Higgs VEVs in MSSM, and $\mu$ is the Higgs mixing 
parameter. 
The soft parameters of \cref{tloop} define our model's parameter set, $\theta=\left\{m_0,m_{1/2},A_0,\tan\beta\right\}$, 
and additionally we consider a set of the most sensitive standard model nuisance parameters, 
$\psi= \left\{m_t,m_b(m_b)^{\overline{\mathrm{MS}}},\alpha_\mathrm{s}(m_Z)^{\overline{\mathrm{MS}}},\alpha_\mathrm{EM}(m_Z)^{\overline{\mathrm{MS}}}\right\}$. 
These together form the basis parameter set: $\Theta = \left\{\theta,\psi\right\}$. Using 
 Bayes's theorem,  the posterior probability density function (PDF) for the
theory described by $\Theta$, which may be mapped to observables, $\xi(\Theta)$ to be compared 
against experimental data, $d$ is given by:
\begin{equation}
  p(\Theta|d) = \frac{p(d|\xi(\Theta))\pi(\Theta)}{p(d)} ~, \label{bayes}
\end{equation}
where $\mathcal{L}\equiv p(d|\xi(\Theta))$ is the likelihood function--the terms of which are described
in \cref{liketable}, $\pi(\Theta)$ is the distribution in $\Theta$ prior to considering 
experimental results, and $\mathcal{Z}\equiv p(d)$ is the Bayesian evidence which can be used in model 
selection. However, in our goal of parameter estimation, it serves only as a normalization factor. We present 
results obtained by considering both the 2D marginalized posterior PDF (where the full N-dimensional posterior 
PDF of \cref{bayes} has been integrated over the other parameters), as well as the profile likelihoods (where the confidence levels are determined 
by comparison to the global best-fit point). (For a more detailed description see the second reference in \cite{multinest}.)

The analysis was done by using {\sc SusyKit}~\cite{susykit}, which 
employs the {\sc MultiNest}~\cite{multinest} package for sampling parameter 
points efficiently, and uses {\sc SoftSUSY}~\cite{Allanach} for spectrum calculation, and {\sc micrOMEGAs}~\cite{belanger} 
to calculate the  relic density as well as for the indirect constraints.
The credible intervals, marginalized posterior PDF's, and profile likelihood distributions 
were calculated using the plotting routines of {\sc SuperBayes}~\cite{superbayes}, which is largely based on 
the tools provided by {\sc CosmoMC}~\cite{cosmomc}.
The constraint from the $g_{\mu}-2$ measurement is not imposed in this analysis and this issue will be 
discussed later in the text.

In our analysis, we took our prior knowledge of the parameters to be either flat linear distributions or flat 
logarithmic distributions, with $m_0\in(0.05,8)\TeV$ (log), $m_{1/2}\in(0.05,5)\TeV$ (log), $A_0\in(-30,30)\TeV$ (linear), 
and $\tan\beta\in(3,60)$ (linear). We have fixed $\sgn(\mu)$ to be positive. The Standard Model nuisance parameters were 
allowed to vary in $2\,\sigma$ windows of their central values, as quoted in \cref{liketable}. Our {\sc MultiNest} sampling 
parameters,   as defined in~\cite{multinest}, were $n_\mathrm{live}=20,000$ and {\tt tol}=0.0001. It has been shown in~\cite{genetic} and the second reference in~\cite{multinest} 
that these parameters are not only sufficient to provide a map of the posterior PDF, but also to find the true best-fit 
point which is essential for the profile likelihood analysis.

{\tiny
\begin{table}[t!]
\begin{center}
{\tiny
\begin{tabular}{|c | c c c |c | c|} \hline
Observable	& Central value	& Exp. Err.	& Th. Err.	& Distribution	& Ref. \\ \hline\hline
\multicolumn{6}{|c|}{SM Nuisance Parameters} \\ \hline\hline
$m_t$           &  $173.5\GeV$	& $1.0\GeV$	& --		& Gaussian 	& \cite{pdgrev}\\
$m_b (m_b)^\msb$& $4.18\GeV$	& $0.03\GeV$	& --		& Gaussian	& \cite{pdgrev} \\
$\alpha_\mathrm{s}(m_Z)^\msb $ & 0.1184 & $7\times10^{-4}$ & --	& Gaussian	& \cite{pdgrev} \\
$1/\alpha_\mathrm{EM}(m_Z)^\msb$ & $127.944$ & 0.014 & --	& Gaussian 	& \cite{pdgrev}\\ \hline\hline
\multicolumn{6}{|c|}{Measured} 	\\ \hline\hline
$\brbsg\times10^{4}$ & 3.21	& 0.33		& 0.21		& Gaussian	& \cite{bphys}\\
$\Omega h^2$	& 0.1126	& 0.0036	& 10\%		& Upper-Gaussian& \cite{WMAP} \\
$m_{h^0}$	& $125.3\GeV$	& $0.6\GeV$	& $1.1\GeV$	& Gaussian 	& \cite{July4}\\ \hline\hline
\multicolumn{6}{|c|}{Limits (95\%~CL)}	\\\hline\hline
$\brbsmm$	& $4.5\times10^{-9}$ & --	& 14\%		& Upper -- Error Fn	& \cite{cmslhcbbsmumu}\\
$m_{h^0}$	& $122.5\GeV$	& --		& --		& Lower -- Step Fn	& \cite{lhc_7tev} \\
$m_{h^0}$	& $129\GeV$	& --		& --		& Upper -- Step Fn	&  \cite{lhc_7tev}\\
$m_{\na}$	& $46\GeV$	& --		& 5\%		& Lower -- Error Fn	& \cite{pdgrev} \\
$m_{\nb}$	& $62.4\GeV$	& --		& 5\%		& Lower -- Error Fn	&\cite{pdgrev} \\
$m_{\nc}$	& $99.9\GeV$	& --		& 5\%		& Lower -- Error Fn	& \cite{pdgrev}\\
$m_{\nd}$	& $116\GeV$	& --		& 5\%		& Lower -- Error Fn	& \cite{pdgrev}\\
$m_{\cha}$	& $94\GeV$ 	& --		& 5\%		& Lower -- Error Fn	& \cite{pdgrev}\\
$m_{\ser}$	& $107\GeV$	& --		& 5\%		& Lower -- Error Fn	&\cite{pdgrev} \\
$m_{\smr}$	& $94\GeV$	& --		& 5\%		& Lower -- Error Fn	&\cite{pdgrev} \\
$m_{\sta}$	& $81.9\GeV$	& --		& 5\%		& Lower -- Error Fn	& \cite{pdgrev}\\
$m_{\ba}$	& $89\GeV$	& --		& 5\%		& Lower -- Error Fn	& \cite{pdgrev}\\
$m_{\ta}$	& $95.7\GeV$	& --		& 5\%		& Lower -- Error Fn	&\cite{pdgrev} \\
$m_{\g}$	& $500\GeV$	& --		& 5\%		& Lower -- Error Fn	& \cite{pdgrev}\\
$m_{\q}$	& $1100\GeV$	& --		& 5\%		& Lower -- Error Fn	&\cite{pdgrev} \\
\hline
\end{tabular}
} 
\caption{\label{liketable} Summary of the observables used to estimate the mSUGRA parameters. Only the 
upper-half of the Gaussian is used in the consideration of $\Omega h^2$, i.e., there is only a 
penalty for values larger than the central value which allows for multicomponent dark matter~\cite{Feldman:2010wy}. The 95\%~CL limits have been evaluated under the 
assumption of only theoretical uncertainty, so the distribution used here is based on the error function, given explicitly in the fourth reference of \cite{superbayes}.  }
\end{center}
\end{table}
}
 
\begin{figure}[t!]
\begin{center}
\includegraphics[scale=0.4]{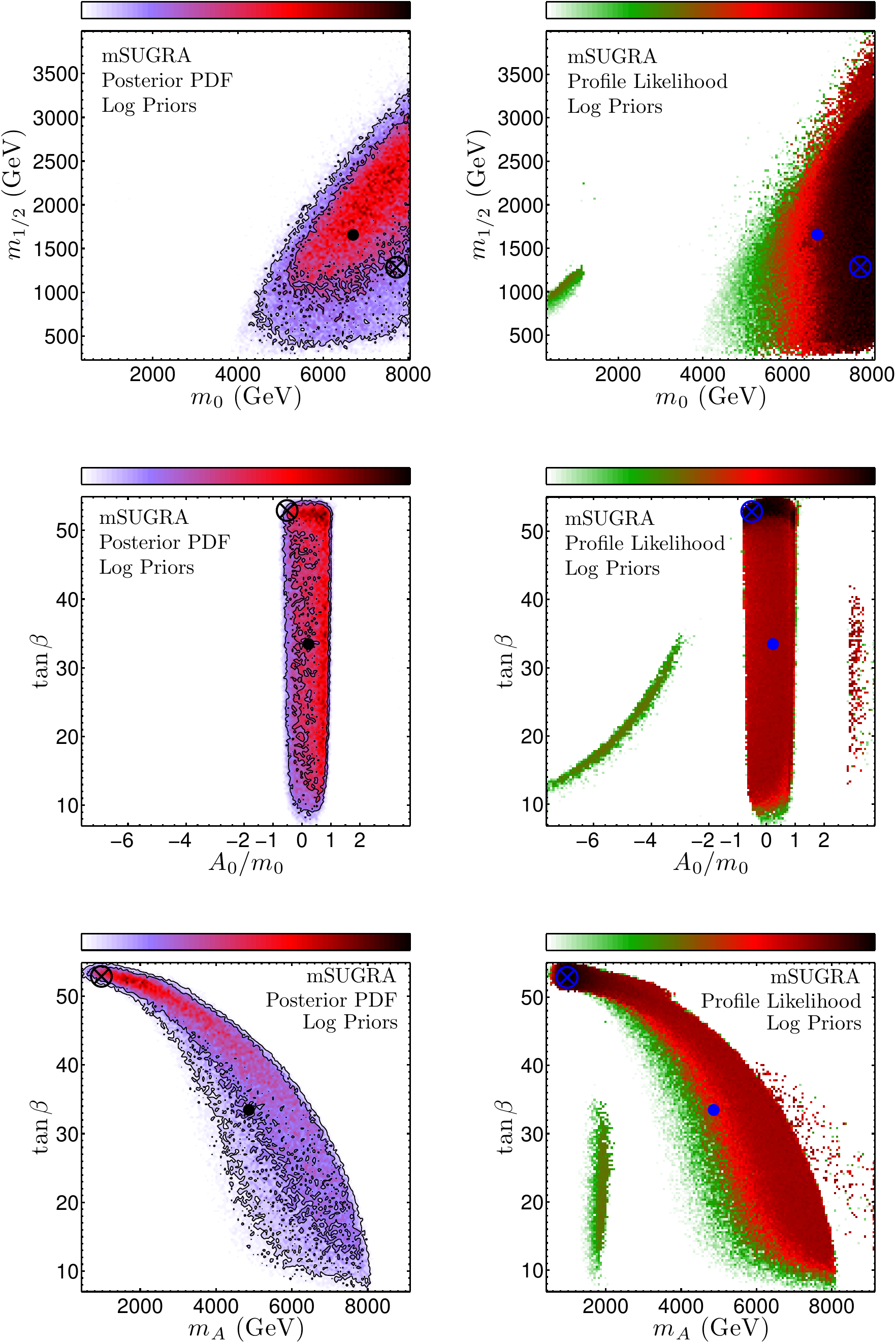}
\caption{
Left panels: plots of the 2D posterior probability densities, $1\,\sigma$ and $2\,\sigma$ contours are also drawn. 
Right panels: plots of the profile likelihoods. 
Top: in the $m_0-\mhf$ plane. Middle: in the $A_0/m_0-\tan\beta$ plane. Bottom: in the $m_A-\tan\beta$ plane.
The posterior mean is marked by a large dot while the best-fit point is shown by a circled `X'. The color bar above the top panel gives the relative likelihood which increases left-to-right.
\label{fig1}}
\end{center}
\end{figure}

In our likelihood analysis we use the CMS result since that result was available earlier~\cite{July4}. We report our fits to the data, including the Higgs mass, in \cref{fig1} in the form 
of 2D posterior PDF maps (left panels) as well as the profile likelihood maps (right panels). The posterior mean is marked with a large dot and the 
global best-fit is marked with a circled `X'. (Note that while the best-fit point is crucial in Frequentist likelihood-ratio tests, it has no significance in 
the Bayesian framework.)
The top panels exhibit the constraint in the $m_0-m_{1/2}$ plane and show that $m_0$ is typically a \tev or larger, while $m_{1/2}$ can lie below $500\GeV$. The middle panels exhibit the constraint
in the $A_0/m_0-\tan\beta$ plane, and here one finds that most of the allowed parameter space lies in the
narrow strip $|A_0/m_0|\leq1$ with a small strip in the range $|A_0/m_0|\in(-2, -6)$.  The bottom panels exhibit the 
constraint in the $m_A-\tan\beta$ plane, and here one finds that the majority of the allowed range of $m_A$ lies above $1\TeV$.
Thus  $m_A\gg m_{h^0}$ for the majority of the parameter space and thus we are in the so-called decoupling limit. 

It was pointed out in~\cite{Akula:2011jx}
 that most of the experimentally consistent parameter space of mSUGRA lies on
 the Hyperbolic Branch (HB)~\cite{Chan:1997bi,Feng:1999mn} of radiative breaking of the electroweak symmetry under the LHC-7 constraints.  The HB region has 
 sub-regions which we may label as Focal Point (HB/FP), Focal Curves (HB/FCi, i=1,2), and Focal Surfaces
 (HB/FS). It was shown in~\cite{Akula:2011aa,Akula:2011jx} that the HB/FP is mostly depleted while the remaining parameter space 
 lies on HB/FCi or HB/FS.  Specifically we note here that the right edge of $A_0/m_0$ in \cref{fig1} is
 $\sim 1$.  The value $|A_0/m_0|=1$  was argued as  string-motivated in~\cite{Feldman:2011ud} and was shown to be the asymptotic limit on the focal curve HB/FC1 in~\cite{Akula:2011jx}.

\begin{figure}[t!]
\begin{center}
\includegraphics[scale=0.6]{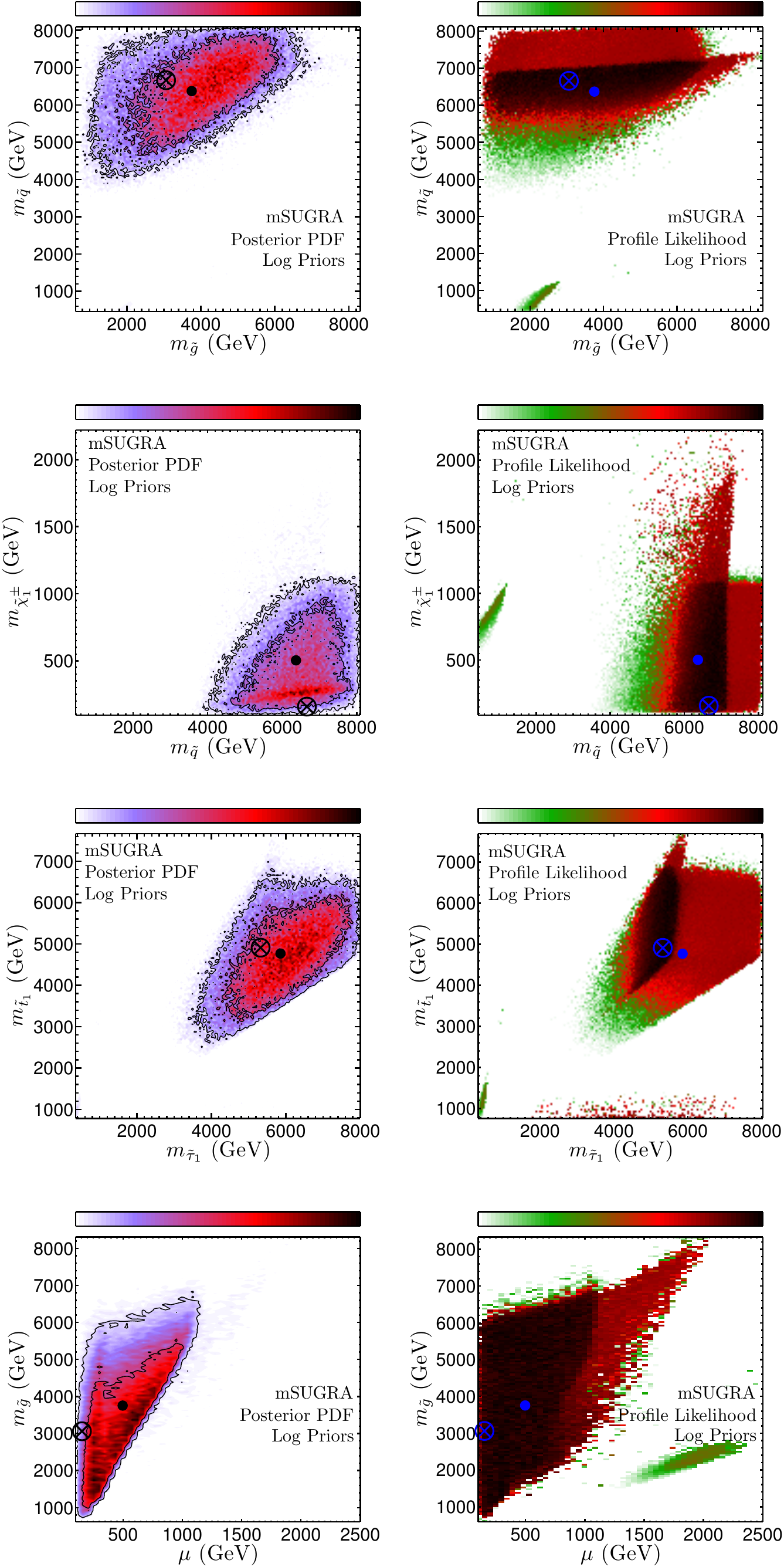}
\caption{
Left panels: plots of the 2D posterior probability densities, $1\,\sigma$ and $2\,\sigma$ contours are also drawn. 
Right panels: plots of the profile likelihoods. 
Top: in the $m_{\tilde g}$--$m_{\tilde q}$ plane. 
Upper-middle: in the $m_{\tilde q}$--$m_{\tilde \chi_1^{\pm}}$ plane. 
Lower-middle: in the $m_{\tilde \tau_1}- m_{\tilde t_1}$ plane.
Bottom: in the $\mu-m_{\tilde g}$ plane.
The posterior mean is marked by a large dot while the best-fit point is shown by a circled `X'. 
The color bar above the top panel gives the relative likelihood which increases left-to-right.
 \label{fig2}}
\end{center}
\end{figure}

In \cref{fig2} we present the 2D posterior PDF's (left panels) and the profile likelihoods (right panels) in the planes of the phenomenologically important sparticle masses. 
The top panels present the results in the gluino--squark mass plane, and indicate that the gluino can be below a \tev.  The second row is plotted in the squark--chargino mass plane and demonstrates that the chargino masses 
are only bounded from below by the direct searches at LEP. 
The next row exhibits our fit in the stau--stop mass plane. Here one finds that the stau and stop masses are typically large except for a small strip where the 
stop mass can lie below a \tev. This is largely to be expected as we rely on a heavy stop to provide a sizable loop correction to the Higgs mass. 
The bottom panels show the analysis in the $\mu-m_{\g}$ plane. One finds that $\mu$ is typically quite light, i.e., $\mu$ can be significantly below $500\GeV$. 

Using the marginalized 1D posterior PDF we are able to set lower limits on the sparticle masses from the $2\,\sigma$ credible regions. We present those limits here: 
$m_{\g}>1.39\TeV$, $m_{\cha}>196\GeV$, $m_{A_0}\sim m_{H_0}\sim m_{H^{\pm}} > 1.3\TeV$, $m_{\ta} > 3.1\TeV$, $m_{\sta}>3.1\TeV$, $m_{\q}>5\TeV$, and $m_{\tilde{\ell}}>4.8\TeV$.
The profile likelihood analysis yields different results. Here, we find the $95\%$ CL sparticle lower limits to be 
$m_{\g}>690\GeV$, $m_{\cha}>95\GeV$, $m_{A_0\sim H_0\sim H^{\pm}} > 540\GeV$, $m_{\ta} > 580\GeV$, $m_{\sta}>310\GeV$, $m_{\q}>1.5\TeV$, and $m_{\tilde{\ell}}>580\GeV$.
We note that as expected the lower limits given by the profile likelihood analysis lie lower than the limits given by the PDF analysis.
The analysis thus indicates  that the light particles in mSUGRA in view of the Higgs mass measurement are 
the neutralino, the chargino, the gluino, the stau and the stop. Among these the most likely candidates for 
discovery  in the next phase of CERN experiment are the gluino, the chargino and the stop.

\begin{figure}[t!]
\begin{center}
\includegraphics[scale=0.4]{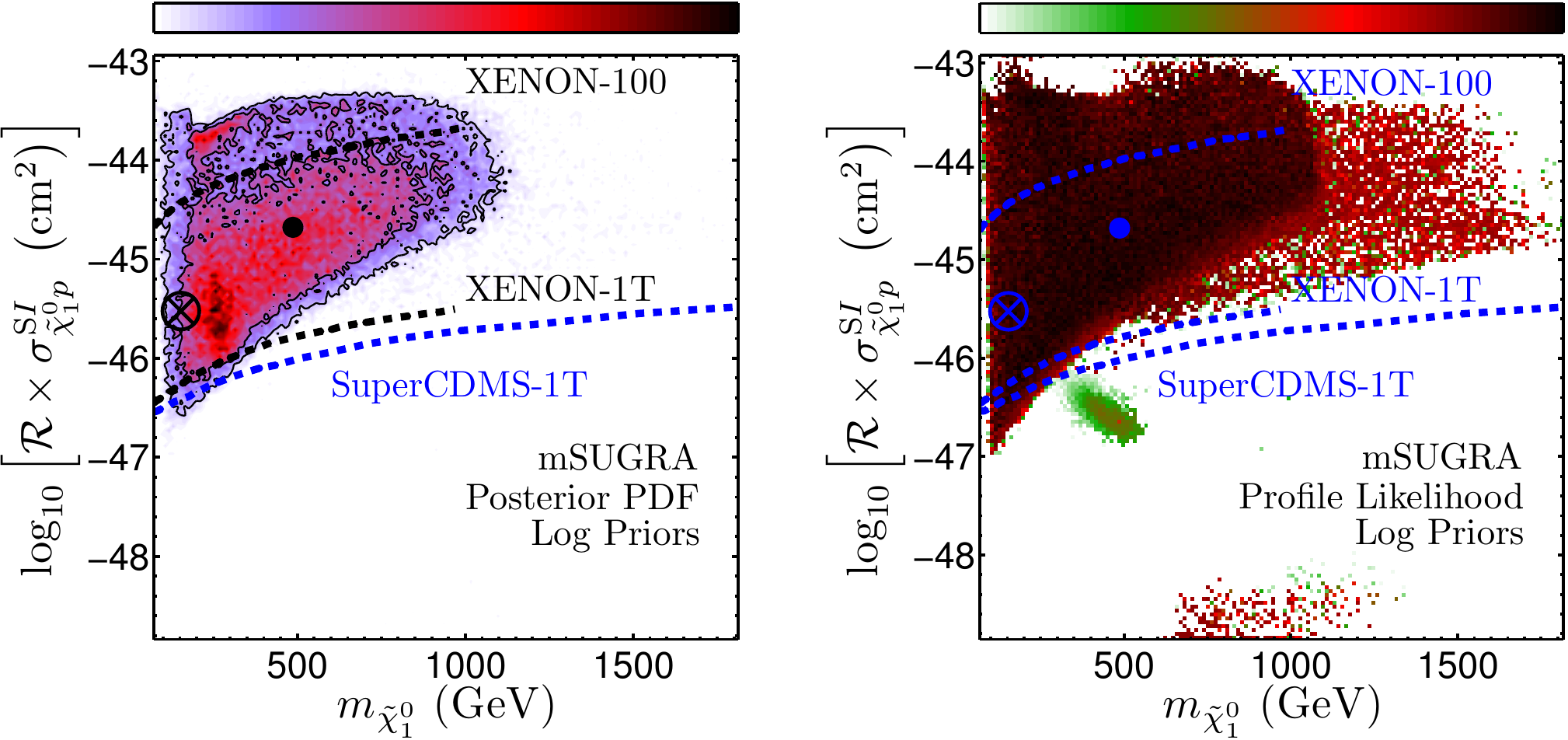}
\caption{(color online) 
Plots of $\mathcal{R}\equiv\left(\Omega h^2\right)/\left(\Omega h^2\right)_{\rm WMAP}$ vs the neutralino mass
$m_{\tilde\chi_1^0}$. The left panel presents the 2D posterior PDF, and the right panel presents the profile 
likelihood.
The analysis shows that virtually all of credible region of mSUGRA will be probed by the 
SuperCDMS~\cite{futureSCDMS}    and      XENON-1T~\cite{futureXENON} experiments.
The color bar above the panels gives the relative likelihood which increases left-to-right.
\label{fig3}}
\end{center}
\end{figure}

{\it $125\GeV$ Higgs boson and dark matter:}
Neutralino-proton spin independent cross section $\sigma_{\na p}^{\rm SI}$ depends sensitively on the Higgs boson
mass (for a discussion see \cite{Akula:2011aa}).  Thus considering the $\sim 125\GeV$ Higgs mass leads to 
a more constrained prediction for dark matter. 
 In  Fig.~\ref{fig3}  we give a plot of $\mathcal{R} \times\sigma_{\na p}^{\rm SI}$ 
 as a function of the lightest neutralino mass $m_{\tilde\chi_1^0}$ where the factor  $\mathcal{R}$ 
 is defined by $\mathcal{R}\equiv\left(\Omega h^2\right)/\left(\Omega h^2\right)_{\rm WMAP}$, and $\left(\Omega h^2\right)_{\rm WMAP}$ 
 is the central value of the WMAP-7 data.
 By only applying a likelihood penalty for points that are above the WMAP-7 limit, we have taken into account the possibility that there 
 may be additional components of dark matter beyond the neutralino~\cite{Feldman:2010wy}. 
   Quite remarkably, the bulk of the credible region of mSUGRA falls essentially exclusively between the current 
   limits on  dark matter by XENON-100~\cite{xenon}  and the projected sensitivity of    
   SuperCDMS~\cite{futureSCDMS}    and  XENON-1T~\cite{futureXENON}. 

We discuss now the constraint from $g_{\mu}-2$. 
In supersymmetric theories, sparticle loops make significant contributions to 
the anomalous magnetic moment of the muon~\cite{Yuan:1984ww} if the relevant sparticles
(charginos, neutralinos, smuons, sneutrinos)  entering the loops 
are relatively light.  The experimental determination of $ \delta a_{\mu} = a_{\mu}^{\rm exp} - a_{\mu}^{\rm SM}$
where $a_{\mu}= (g_{\mu}-2)/2$, depends sensitively on  the hadronic correction to the standard model value. 
There are two main procedures for the estimation of the hadronic correction, which are either using the $e^+e^-$ annihilation cross section or 
from $\tau$ decay.  The result using the $e^+ e^-$ annihilation gives $\delta a_{\mu}= (28.7\pm 8.0)\times 10^{-10} ~(3.6\,\sigma)$
while for $\tau$-based hadronic contributions one has
$\delta a_{\mu}= (19.5\pm 8.3)\times 10^{-10} ~(2.4\,\sigma)$~\cite{hoecker}. In any case, within the universal 
soft SUSY-breaking paradigm there would be tension between the $g_{\mu}-2$ result (specifically  the
one using $e^+e^-$ annihilation cross section) and the $125\GeV$ Higgs boson mass since the
$m_0$ scale is rather  high. If the $g_{\mu}-2$ results stay,  then there are at least two avenues
open to make compatible the $g_{\mu}-2$ results and the Higgs boson mass.
The first possibility is that we stay within the universal soft breaking paradigm and additional
contributions to the Higgs mass  arise due
to the  presence of extra matter which can generate new loop corrections to the Higgs mass, or from  extra gauge groups 
under which the Higgs is charged yielding corrections to the Higgs mass through $D$-terms.
 Alternatively, one  could give up universality of soft parameters and consider non-universal or flavored SUGRA 
models~\cite{Nath:1997qm}. For instance,  to satisfy the $g_{\mu}-2$ constraint  one may consider  the soft  scalar mass for the first two generations 
much smaller than for the third generation, or the sleptons being lighter than the squarks. These possibilities require  further investigation. \\  

{\it Conclusion:} 
In this work we have analyzed the 
implications of the Higgs boson discovery at CERN for supersymmetry. Specifically we analyzed the mSUGRA model to delineate
constraints on soft parameters and identified the light particles that are prime candidates for discovery in the next phase 
of runs at the LHC. The analysis presented here explains why supersymmetric dark matter has not been seen thus  far 
since essentially all of the parameter space lies below the current sensitivity of dark matter experiments due to the high Higgs mass.
The analysis also points to excellent prospects for the discovery of dark matter at SuperCDMS and XENON-1T as well as the possibility 
of light neutralinos, charginos and gluinos, and possibly light stops and staus.

\begin{acknowledgments}
One of us (S.A.) thanks TASI-2012 where part of this work was done. 
This research is supported in part by NSF Grants Nos. PHY-0757959, and PHY-0969739, and 
used resources of the National Energy Research Scientific Computing Center, which is
supported by the Office of Science of the U.S. Department of Energy
under Contract No. DE-AC02-05CH11231.
\end{acknowledgments}

\end{document}